
1 Postscript figure is found at XXX.

%
%
  \font\twelverm=cmr10 scaled 1200       \font\twelvei=cmmi10 scaled 1200
  \font\twelvesy=cmsy10 scaled 1200      \font\twelveex=cmex10 scaled 1200
  \font\twelvebf=cmbx10 scaled 1200      \font\twelvesl=cmsl10 scaled 1200
  \font\twelvett=cmtt10 scaled 1200      \font\twelveit=cmti10 scaled 1200
  \font\twelvemib=cmmib10 scaled 1200
  \font\elevenmib=cmmib10 scaled 1095
  \font\tenmib=cmmib10
  \font\eightmib=cmmib10 scaled 800
  
\font\elevenrm=cmr10 scaled 1095    \font\eleveni=cmmi10 scaled 1095
\font\elevensy=cmsy10 scaled 1095

%
%

\font\seventeeni=cmmi10 scaled \magstep3

\font\seventeensy=cmsy10 scaled \magstep3

\font\seventeenmib=cmmib10 scaled \magstep3

\newfam\cpfam%



\skewchar\eleveni='177   \skewchar\elevensy='60
\skewchar\elevenmib='177  \skewchar\seventeensy='60
\skewchar\seventeenmib='177
\skewchar\seventeeni='177

\newfam\mibfam%


  \skewchar\twelvei='177   \skewchar\twelvesy='60
  \skewchar\twelvemib='177
%
%
\def\twelvepoint{\normalbaselineskip=12.4pt
  \abovedisplayskip 12.4pt plus 3pt minus 9pt
  \belowdisplayskip 12.4pt plus 3pt minus 9pt
  \abovedisplayshortskip 0pt plus 3pt
  \belowdisplayshortskip 7.2pt plus 3pt minus 4pt
  \smallskipamount=3.6pt plus 1.2pt minus 1.2pt
  \medskipamount=7.2pt plus 2.4pt minus 2.4pt
  \bigskipamount=14.4pt plus 4.8pt minus 4.8pt
  \def\rm{\fam0\twelverm}          \def\it{\fam\itfam\twelveit}%
  \def\sl{\fam\slfam\twelvesl}     \def\bf{\fam\bffam\twelvebf}%
  \def\mit{\fam 1}                 \def\cal{\fam 2}%
  \def\tt{\twelvett}%
  \def\mib{\fam\mibfam\twelvemib}%

  \textfont0=\twelverm   \scriptfont0=\tenrm     \scriptscriptfont0=\sevenrm
  \textfont1=\twelvei    \scriptfont1=\teni      \scriptscriptfont1=\seveni
  \textfont2=\twelvesy   \scriptfont2=\tensy     \scriptscriptfont2=\sevensy
  \textfont3=\twelveex   \scriptfont3=\twelveex  \scriptscriptfont3=\twelveex
  \textfont\itfam=\twelveit
  \textfont\slfam=\twelvesl
  \textfont\bffam=\twelvebf
  \textfont\mibfam=\twelvemib       \scriptfont\mibfam=\tenmib
                                             \scriptscriptfont\mibfam=\eightmib

  \def\xrm{\textfont0=\twelverm\scriptfont0=\tenrm
      \scriptscriptfont0=\sevenrm\rm}
\normalbaselines\rm}


\mathchardef\alpha="710B
\mathchardef\beta="710C
\mathchardef\gamma="710D
\mathchardef\delta="710E
\mathchardef\epsilon="710F
\mathchardef\zeta="7110
\mathchardef\eta="7111
\mathchardef\theta="7112
\mathchardef\kappa="7114
\mathchardef\lambda="7115
\mathchardef\mu="7116
\mathchardef\nu="7117
\mathchardef\xi="7118
\mathchardef\pi="7119
\mathchardef\rho="711A
\mathchardef\sigma="711B
\mathchardef\tau="711C
\mathchardef\phi="711E
\mathchardef\chi="711F
\mathchardef\psi="7120
\mathchardef\omega="7121
\mathchardef\varepsilon="7122
\mathchardef\vartheta="7123
\mathchardef\varrho="7125
\mathchardef\varphi="7127

\def\physgreek{
\mathchardef\Gamma="7100
\mathchardef\Delta="7101
\mathchardef\Theta="7102
\mathchardef\Lambda="7103
\mathchardef\Xi="7104
\mathchardef\Pi="7105
\mathchardef\Sigma="7106
\mathchardef\Upsilon="7107
\mathchardef\Phi="7108
\mathchardef\Psi="7109
\mathchardef\Omega="710A}


\def\beginlinemode{\endmode
  \begingroup\parskip=0pt \obeylines\def\\{\par}\def\endmode{\par\endgroup}}
\def\beginparmode{\endmode
  \begingroup \def\endmode{\par\endgroup}}
\let\endmode=\par
{\obeylines\gdef\
{}}
\def\singlespace{\baselineskip=\normalbaselineskip}

\def\oneandahalfspace{\baselineskip=\normalbaselineskip
  \multiply\baselineskip by 3 \divide\baselineskip by 2}
\def\doublespace{\baselineskip=\normalbaselineskip \multiply\baselineskip by 2}

\nopagenumbers
\newcount\firstpageno
\firstpageno=2
\headline={\ifnum\pageno<\firstpageno{\hfil}\else{\hfil\elevenrm\folio\hfil}\fi}
\let\rawfootnote=\footnote             
\def\footnote#1#2{{\singlespace\parindent=0pt
\rawfootnote{#1}{#2}}}
\def\raggedcenter{\leftskip=4em plus 12em \rightskip=\leftskip
  \parindent=0pt \parfillskip=0pt \spaceskip=.3333em \xspaceskip=.5em
  \pretolerance=9999 \tolerance=9999
  \hyphenpenalty=9999 \exhyphenpenalty=9999 }
\def\dateline{\rightline{\ifcase\month\or
  January\or February\or March\or April\or May\or June\or
  July\or August\or September\or October\or November\or December\fi
  \space\number\year}}
\def\received{\vskip 3pt plus 0.2fill
 \centerline{\sl (Received\space\ifcase\month\or
  January\or February\or March\or April\or May\or June\or
  July\or August\or September\or October\or November\or December\fi
  \qquad, \number\year)}}


\hsize=6.5truein
\hoffset=0.0truein
\vsize=8.9truein
\voffset=0truein
\hfuzz=0.1pt
\vfuzz=0.1pt
\parskip=\medskipamount
\overfullrule=0pt      



\def\title                     
  {\null\vskip 3pt plus 0.1fill
   \beginlinemode \doublespace \raggedcenter \bf}

\def\author                    
  {\vskip 6pt plus 0.2fill \beginlinemode
   \singlespace \raggedcenter}

\def\affil        
  {\vskip 6pt plus 0.1fill \beginlinemode
   \oneandahalfspace \raggedcenter \it}

\def\abstract                  
  {\vskip 6pt plus 0.3fill \beginparmode
   \doublespace \narrower }

\def\summary                   
  {\vskip 3pt plus 0.3fill \beginparmode
   \doublespace \narrower SUMMARY: }

\def\pacs#1
  {\vskip 3pt plus 0.2fill PACS numbers: #1}

\def\endtitlepage              
  {\endpage                    
   \body}

\def\body                      
  {\beginparmode}              

\def\head#1{                   
  \filbreak\vskip 0.5truein    
  {\immediate\write16{#1}
   \raggedcenter \uppercase{#1}\par}
   \nobreak\vskip 0.25truein\nobreak}

%
%

%
\def\inlinerefs{
  \gdef\refto##1{ [##1]}                
\gdef\refis##1{\indent\hbox to 0pt{\hss##1.~}} 
\gdef\journal##1, ##2, ##3, 1##4##5##6{ 
    {\sl ##1~}{\bf ##2}, ##3 (1##4##5##6)}}    
\def\keywords#1
  {\vskip 3pt plus 0.2fill Keywords: #1}
\gdef\figis#1{\indent\hbox to 0pt{\hss#1.~}} 

\def\figurecaptions     
  {\head{Figure Captions}    
   \beginparmode
   \interlinepenalty=10000
   \frenchspacing \parindent=0pt \leftskip=1truecm
   \parskip=8pt plus 3pt \everypar{\hangindent=\parindent}}

%
%
\def\refto#1{$^{#1}$}          

\def\references       
  {\head{References}           
   \beginparmode
   \frenchspacing \parindent=0pt \leftskip=1truecm
   \interlinepenalty=10000
   \parskip=8pt plus 3pt \everypar{\hangindent=\parindent}}

\gdef\refis#1{\indent\hbox to 0pt{\hss#1.~}} 

\gdef\journal#1, #2, #3, 1#4#5#6{              
    {\sl #1~}{\bf #2}, #3 (1#4#5#6)}          

\def\refstylenp{               
  \gdef\refto##1{ [##1]}                               
  \gdef\refis##1{\indent\hbox to 0pt{\hss##1)~}}      
  \gdef\journal##1, ##2, ##3, ##4 {                    
     {\sl ##1~}{\bf ##2~}(##3) ##4 }}

\def\refstyleprnp{             
  \gdef\refto##1{ [##1]}                               
  \gdef\refis##1{\indent\hbox to 0pt{\hss##1)~}}      
  \gdef\journal##1, ##2, ##3, 1##4##5##6{              
    {\sl ##1~}{\bf ##2~}(1##4##5##6) ##3}}

\def\prl{\journal Phys. Rev. Lett., }

\def\endreferences{\body}

%
%

\def\endfigurecaptions{\body}

\def\endpage                   
  {\vfill\eject}

\def\endpaper                  
  {\endmode\vfill\supereject}

\def\endit
  {\endpaper\end}


\def\ref#1{Ref.[#1]}                   
\def\Ref#1{Ref.[#1]}                   

\def\Equation#1{Equation [#1]}         
\def\Equations#1{Equations [#1]}       
\def\Eq#1{Eq. (#1)}                     
\def\eq#1{Eq. (#1)}                     
\def\Eqs#1{Eqs. (#1)}                   
\def\eqs#1{Eqs. (#1)}                   
\def\frac#1#2{{\textstyle{{\strut #1} \over{\strut #2}}}}

\def\sla{\raise.15ex\hbox{$/$}\kern-.57em}
\def\leaderfill{\leaders\hbox to 1em{\hss.\hss}\hfill}
\def\twiddle{\lower.9ex\rlap{$\kern-.1em\scriptstyle\sim$}}
\def\bigtwiddle{\lower1.ex\rlap{$\sim$}}
\def\gtwid{\mathrel{\raise.3ex\hbox{$>$\kern-.75em\lower1ex\hbox{$\sim$}}}}
\def\ltwid{\mathrel{\raise.3ex\hbox{$<$\kern-.75em\lower1ex\hbox{$\sim$}}}}
\def\square{\kern1pt\vbox{\hrule height 1.2pt\hbox{\vrule width 1.2pt\hskip 3pt
   \vbox{\vskip 6pt}\hskip 3pt\vrule width 0.6pt}\hrule height 0.6pt}\kern1pt}

%

%

%

%
\physgreek
%

\def\dsl{\raise.15ex\hbox{$/$}\kern-.57em\hbox{$\partial$}}
\def\nsl{\raise.15ex\hbox{$/$}\kern-.57em\hbox{$\nabla$}}
\def\gtwid{\,{\raise.3ex\hbox{$>$\kern-.75em\lower1ex\hbox{$\sim$}}}\,}
\def\ltwid{\,{\raise.3ex\hbox{$<$\kern-.75em\lower1ex\hbox{$\sim$}}}\,}
\def\undr{\raise.3ex\hbox{$\sim$\kern-.75em\lower1ex\hbox{$|\vec
x|\to\infty$}}}

\def\[{\left [}
\def\]{\right ]}
\def\({\left (}
\def\){\right )}







\def\and{a^{\phantom\dagger}}

%
\def\id{\raise.72ex\hbox{$-$}\kern-.85em\hbox{$d$}\,}

\catcode`@=11
\newcount\r@fcount \r@fcount=0
\newcount\r@fcurr
\immediate\newwrite\reffile
\newif\ifr@ffile\r@ffilefalse
\def\w@rnwrite#1{\ifr@ffile\immediate\write\reffile{#1}\fi\message{#1}}

\def\writer@f#1>>{}
\def\referencefile{
  \r@ffiletrue\immediate\openout\reffile=\jobname.ref%
  \def\writer@f##1>>{\ifr@ffile\immediate\write\reffile%
    {\noexpand\refis{##1} = \csname r@fnum##1\endcsname = %
     \expandafter\expandafter\expandafter\strip@t\expandafter%
     \meaning\csname r@ftext\csname r@fnum##1\endcsname\endcsname}\fi}%
  \def\strip@t##1>>{}}

\def\citeall#1{\xdef#1##1{#1{\noexpand\cite{##1}}}}
\def\cite#1{\each@rg\citer@nge{#1}}	

\def\each@rg#1#2{{\let\thecsname=#1\expandafter\first@rg#2,\end,}}
\def\first@rg#1,{\thecsname{#1}\apply@rg}	
\def\apply@rg#1,{\ifx\end#1\let\next=\relax
\else,\thecsname{#1}\let\next=\apply@rg\fi\next}

\def\citer@nge#1{\citedor@nge#1-\end-}	
\def\citer@ngeat#1\end-{#1}
\def\citedor@nge#1-#2-{\ifx\end#2\r@featspace#1 
  \else\citel@@p{#1}{#2}\citer@ngeat\fi}	
\def\citel@@p#1#2{\ifnum#1>#2{\errmessage{Reference range #1-#2\space is bad.}%
    \errhelp{If you cite a series of references by the notation M-N, then M and
    N must be integers, and N must be greater than or equal to M.}}\else%
 {\count0=#1\count1=#2\advance\count1
by1\relax\expandafter\r@fcite\the\count0,%
  \loop\advance\count0 by1\relax
    \ifnum\count0<\count1,\expandafter\r@fcite\the\count0,%
  \repeat}\fi}

\def\r@featspace#1#2 {\r@fcite#1#2,}	
\def\r@fcite#1,{\ifuncit@d{#1}
    \newr@f{#1}%
    \expandafter\gdef\csname r@ftext\number\r@fcount\endcsname%
                     {\message{Reference #1 to be supplied.}%
                      \writer@f#1>>#1 to be supplied.\par}%
 \fi%
 \csname r@fnum#1\endcsname}
\def\ifuncit@d#1{\expandafter\ifx\csname r@fnum#1\endcsname\relax}%
\def\newr@f#1{\global\advance\r@fcount by1%
    \expandafter\xdef\csname r@fnum#1\endcsname{\number\r@fcount}}

\let\r@fis=\refis			
\def\refis#1#2#3\par{\ifuncit@d{#1}
   \newr@f{#1}%
   \w@rnwrite{Reference #1=\number\r@fcount\space is not cited up to now.}\fi%
  \expandafter\gdef\csname r@ftext\csname r@fnum#1\endcsname\endcsname%
  {\writer@f#1>>#2#3\par}}

\def\ignoreuncited{
   \def\refis##1##2##3\par{\ifuncit@d{##1}%
     \else\expandafter\gdef\csname r@ftext\csname
r@fnum##1\endcsname\endcsname%
     {\writer@f##1>>##2##3\par}\fi}}

\def\r@ferr{\endreferences\errmessage{I was expecting to see
\noexpand\endreferences before now;  I have inserted it here.}}
\let\r@ferences=\references
\def\references{\r@ferences\def\endmode{\r@ferr\par\endgroup}}

\let\endr@ferences=\endreferences
\def\endreferences{\r@fcurr=0
  {\loop\ifnum\r@fcurr<\r@fcount
    \advance\r@fcurr by 1\relax\expandafter\r@fis\expandafter{\number\r@fcurr}%
    \csname r@ftext\number\r@fcurr\endcsname%
  \repeat}\gdef\r@ferr{}\endr@ferences}


\let\r@fend=\endpaper\gdef\endpaper{\ifr@ffile
\immediate\write16{Cross References written on []\jobname.REF.}\fi\r@fend}

\catcode`@=12

\citeall\refto		
\citeall\ref		%
\citeall\Ref		%

\catcode`@=11
\newcount\tagnumber\tagnumber=0

\immediate\newwrite\eqnfile
\newif\if@qnfile\@qnfilefalse
\def\write@qn#1{}
\def\writenew@qn#1{}
\def\w@rnwrite#1{\write@qn{#1}\message{#1}}
\def\@rrwrite#1{\write@qn{#1}\errmessage{#1}}

\def\taghead#1{\gdef\t@ghead{#1}\global\tagnumber=0}
\def\t@ghead{}

\expandafter\def\csname @qnnum-3\endcsname
  {{\t@ghead\advance\tagnumber by -3\relax\number\tagnumber}}
\expandafter\def\csname @qnnum-2\endcsname
  {{\t@ghead\advance\tagnumber by -2\relax\number\tagnumber}}
\expandafter\def\csname @qnnum-1\endcsname
  {{\t@ghead\advance\tagnumber by -1\relax\number\tagnumber}}
\expandafter\def\csname @qnnum0\endcsname
  {\t@ghead\number\tagnumber}
\expandafter\def\csname @qnnum+1\endcsname
  {{\t@ghead\advance\tagnumber by 1\relax\number\tagnumber}}
\expandafter\def\csname @qnnum+2\endcsname
  {{\t@ghead\advance\tagnumber by 2\relax\number\tagnumber}}
\expandafter\def\csname @qnnum+3\endcsname
  {{\t@ghead\advance\tagnumber by 3\relax\number\tagnumber}}

\def\equationfile{%
  \@qnfiletrue\immediate\openout\eqnfile=\jobname.eqn%
  \def\write@qn##1{\if@qnfile\immediate\write\eqnfile{##1}\fi}
  \def\writenew@qn##1{\if@qnfile\immediate\write\eqnfile
    {\noexpand\tag{##1} = (\t@ghead\number\tagnumber)}\fi}
}

\def\callall#1{\xdef#1##1{#1{\noexpand\call{##1}}}}
\def\call#1{\each@rg\callr@nge{#1}}

\def\each@rg#1#2{{\let\thecsname=#1\expandafter\first@rg#2,\end,}}
\def\first@rg#1,{\thecsname{#1}\apply@rg}
\def\apply@rg#1,{\ifx\end#1\let\next=\relax%
\else,\thecsname{#1}\let\next=\apply@rg\fi\next}

\def\callr@nge#1{\calldor@nge#1-\end-}
\def\callr@ngeat#1\end-{#1}
\def\calldor@nge#1-#2-{\ifx\end#2\@qneatspace#1 %
  \else\calll@@p{#1}{#2}\callr@ngeat\fi}
\def\calll@@p#1#2{\ifnum#1>#2{\@rrwrite{Equation range #1-#2\space is bad.}
\errhelp{If you call a series of equations by the notation M-N, then M and
N must be integers, and N must be greater than or equal to M.}}\else%
 {\count0=#1\count1=#2\advance\count1
 by1\relax\expandafter\@qncall\the\count0,%
  \loop\advance\count0 by1\relax%
    \ifnum\count0<\count1,\expandafter\@qncall\the\count0,%
  \repeat}\fi}

\def\@qneatspace#1#2 {\@qncall#1#2,}
\def\@qncall#1,{\ifunc@lled{#1}{\def\next{#1}\ifx\next\empty\else
  \w@rnwrite{Equation number \noexpand\(>>#1<<) has not been defined yet.}
  >>#1<<\fi}\else\csname @qnnum#1\endcsname\fi}

\let\eqnono=\eqno
\def\eqno(#1){\tag#1}
\def\tag#1$${\eqnono(\displayt@g#1 )$$}

\def\aligntag#1\endaligntag
  $${\gdef\tag##1\\{&(##1 )\cr}\eqalignno{#1\\}$$
  \gdef\tag##1$${\eqnono(\displayt@g##1 )$$}}

\def\eqalignno#1{\displ@y \tabskip\centering
  \halign to\displaywidth{\hfil$\displaystyle{##}$\tabskip\z@skip
    &$\displaystyle{{}##}$\hfil\tabskip\centering
    &\llap{$\displayt@gpar##$}\tabskip\z@skip\crcr
    #1\crcr}}

\def\displayt@gpar(#1){(\displayt@g#1 )}

\def\displayt@g#1 {\rm\ifunc@lled{#1}\global\advance\tagnumber by1
        {\def\next{#1}\ifx\next\empty\else\expandafter
        \xdef\csname @qnnum#1\endcsname{\t@ghead\number\tagnumber}\fi}%
  \writenew@qn{#1}\t@ghead\number\tagnumber\else
        {\edef\next{\t@ghead\number\tagnumber}%
        \expandafter\ifx\csname @qnnum#1\endcsname\next\else
        \w@rnwrite{Equation \noexpand\tag{#1} is a duplicate number.}\fi}%
  \csname @qnnum#1\endcsname\fi}

\def\ifunc@lled#1{\expandafter\ifx\csname @qnnum#1\endcsname\relax}

\let\@qnend=\end\gdef\end{\if@qnfile
\immediate\write16{Equation numbers written on []\jobname.EQN.}\fi\@qnend}

\catcode`@=12
\callall\Equation
\callall\Equations
\callall\Eq
\callall\eq
\callall\Eqs
\callall\eqs


\referencefile

\twelvepoint\doublespace

\title{Symmetry of the Gap in Bi2212 from Photoemission
Spectroscopy}

\author{M. R. Norman}
\affil
Materials Science Division
Argonne National Laboratory
Argonne, IL  60439

\body

\noindent PACS numbers:  74.70.Vy, 79.60.-i

\bigskip

In a recent Letter, Shen et al\refto{Shen} have detected a large anisotropy
of the superconducting gap in $Bi_2Sr_2CaCu_2O_8$ (Bi2212), consistent
with d-wave symmetry, from photoemission spectroscopy.  Moreover, they
claim that the change in their spectra as a function of aging is also
consistent
with such an intrepretation.  In this Comment, I show that the latter
statement is not entirely correct, in that the data as a function of aging are
inconsistent with a d-wave gap but are
consistent with an anisotropic s-wave gap.

In Ref. 1, the data show that the gap is close to zero along the
(1,1) direction and maximum along the (1,0) direction, just as
expected for a gap with $d_{x^2-y^2}$ symmetry.  As the sample ages,
the gap becomes non-zero along (1,1).  This is attributed by them to
poorer k resolution due to impurity scattering.  This brings up the interesting
question of what one theoretically expects for the excitation gap as
a function of impurities.  I do this for two cases where a spherical Fermi
cylinder is assumed for simplicity's sake.  For the d-wave case, the gap
is assumed to be $\Delta cos(2\phi)$ and for the anisotropic s-wave
case,
$\Delta |cos(2\phi)|$.  Thus, the modulus of the gap is the same, the
only difference being the symmetry of the gap under rotations.

The self-energy equations at T=0 in the Born approximation
are\refto{Abrikosov,Ueda}
$$
\tilde \omega = \omega + {4\Gamma \over \pi} \int_0^{{\pi \over 4}}
d\phi'{\tilde \omega
\over \sqrt{\tilde \Delta(\phi')^2 - \tilde \omega^2}}
\eqno(1)
$$
$$
\tilde \Delta(\phi) = \Delta(\phi) + {4\Gamma \over \pi} \int_0^{{\pi \over 4}}
d\phi'
{\tilde \Delta(\phi')
\over \sqrt{\tilde \Delta(\phi')^2 - \tilde \omega^2}}
\eqno(2)
$$
where $\Gamma$ is the strength of the impurity scattering.
These solutions are then used
to determine the angle-resolved density of states
$$
N(\phi,\omega) = N_0 Im({\tilde \omega \over \sqrt{\tilde \Delta(\phi)^2 -
\tilde \omega^2}})
\eqno(3)
$$
where $N_0$ is the density of states in the normal state.
The excitation gap is where N first becomes non-zero.  For the d-wave
case, this is zero for all $\phi$ as soon as $\Gamma$ is non-zero.
Instead, the excitation gap employed here is the same as that defined
in Ref. 1, that is, where the value of $N/N_0$ first becomes equal to 1/2.
These results are summarized in the Figure.  As can be seen, the d-wave case
is qualitatively inconsistent with the data, since it becomes rapidly gapless
as $\Gamma$ increases.  In contrast, the
anisotropic s-wave case gives a behavior consistent with the data, with the
gap becoming isotropic (equal to $\Delta/2$) in the large $\Gamma$ limit.

A final remark is that the gap equations have not been resolved ($\Delta$
was treated as fixed).  One can easily calculate the reduction of $T_c$ with
$\Gamma$
$$
ln({T_{c0} \over T_c}) = c(\psi({1 \over 2} + {\Gamma \over 2 \pi T_c}) -
\psi({1 \over 2}))
\eqno(4)
$$
where $T_{c0}$ is $T_c$ for the pure system and c is 1 for the d-wave case
and $1 - 8/\pi^2 \sim 0.19$ for the s-wave case.  Similarly,
$\Delta$ decreases as $\Gamma$ increases with the effect being much larger
in the d-wave case.  This makes the d-wave case even more inconsistent with
the data.

This work was supported by U.S. Department of Energy, Office of Basic Energy
Sciences, under Contract No. W-31-109-ENG-38.

\bigskip

\references

\refis{Shen} Z.-X. Shen et al, \prl 70, 1553, 1993.

\refis{Abrikosov} A. A. Abrikosov and L. P. Gor'kov, JETP {\bf 12}, 1243
(1961).

\refis{Ueda} K. Ueda and T. M. Rice, in Theory of Heavy Fermions and Valence
Fluctuations, eds. T. Kasuya and T. Saso (Springer-Verlag, New York, 1985),
p. 267.

\endreferences

\bigskip

\figurecaptions

\figis{1} Excitation gap as a function of position on the Fermi surface
for various impurity concentrations for d-wave case and anisotropic
s-wave case.  The lines are labeled by values of $\Gamma/\Delta$.
The solid points are data from Ref. 1.

\endfigurecaptions

\vfill\eject

\endit

XXX

/md 194 dict def md begin
/currentpacking where {pop /sc_oldpacking currentpacking def true setpacking}if
/bd{bind def}bind def
/xdf{exch def}bd
/xs{exch store}bd
/ld{load def}bd
/Z{0 def}bd
/T/true
/F/false
/:L/lineto
/lw/setlinewidth
/:M/moveto
/rl/rlineto
/rm/rmoveto
/:C/curveto
/:T/translate
/:K/closepath
/:mf/makefont
/gS/gsave
/gR/grestore
/np/newpath
14{ld}repeat
/$m matrix def
/av 81 def
/por true def
/normland false def
/psb-nosave{}bd
/pse-nosave{}bd
/us Z
/psb{/us save store}bd
/pse{us restore}bd
/level2
/languagelevel where
{
pop languagelevel 2 ge
}{
false
}ifelse
def
/featurecleanup
{
stopped
cleartomark
countdictstack exch sub dup 0 gt
{
{end}repeat
}{
pop
}ifelse
}bd
/noload Z
/startnoload
{
{/noload save store}if
}bd
/endnoload
{
{noload restore}if
}bd
level2 startnoload
/setjob
{
statusdict/jobname 3 -1 roll put
}bd
/setcopies
{
userdict/#copies 3 -1 roll put
}bd
level2 endnoload level2 not startnoload
/setjob
{
1 dict begin/JobName xdf currentdict end setuserparams
}bd
/setcopies
{
1 dict begin/NumCopies xdf currentdict end setpagedevice
}bd
level2 not endnoload
/pm Z
/mT Z
/sD Z
/realshowpage Z
/initializepage
{
/pm save store mT concat
}bd
/endp
{
pm restore showpage
}def
/$c/DeviceRGB def
/rectclip where
{
pop/rC/rectclip ld
}{
/rC
{
np 4 2 roll
:M
1 index 0 rl
0 exch rl
neg 0 rl
:K
clip np
}bd
}ifelse
/rectfill where
{
pop/rF/rectfill ld
}{
/rF
{
gS
np
4 2 roll
:M
1 index 0 rl
0 exch rl
neg 0 rl
fill
gR
}bd
}ifelse
/rectstroke where
{
pop/rS/rectstroke ld
}{
/rS
{
gS
np
4 2 roll
:M
1 index 0 rl
0 exch rl
neg 0 rl
:K
stroke
gR
}bd
}ifelse
/G/setgray ld
/:F/setrgbcolor ld
level2 startnoload
/$i false def
/flipinvert
statusdict begin
version cvr 47.0 lt
end
def
/iw Z
/ih Z
/im_save Z
/setupimageproc Z
/polarity Z
/smoothflag Z
/$z Z
/bpc Z
/smooth_moredata Z
/datatype Z
/:f
{
/im_save save store
/datatype xs
$i flipinvert
and
xor
/polarity xs
/smoothflag xs
:T
scale
/$z exch string store
/bpc xs
/ih xs
/iw xs
/smoothflag
smoothflag
bpc 1 eq and
smoothflag and
userdict/sc_smooth known and
vmstatus pop exch pop iw 3 mul sub 1000 gt and
iw 4 mul 7 add 8 idiv 4 mul 65535 le and
store
smoothflag{
iw
ih
$z
iw 7 add 8 idiv 4 mul string
iw 4 mul 7 add 8 idiv 4 mul string
true
false
sc_initsmooth
/iw iw 4 mul store
/ih ih 4 mul store
}if
/setupimageproc datatype 0 eq datatype 1 eq or{
smoothflag{
{
[
/smooth_moredata cvx[
currentfile
$z
{readstring readhexstring}datatype get
/pop cvx
]cvx[
$z
]cvx/ifelse cvx
/sc_smooth cvx
/smooth_moredata/exch cvx/store cvx
]cvx bind
/smooth_moredata true store
dup exec pop dup exec pop
}
}{
{
[
currentfile
$z
{readstring readhexstring}datatype get
/pop cvx
]cvx bind
}
}ifelse
}{
(error, can't use level2 data acquisition procs for level1)print flush stop
}ifelse
store
}bd
/:j{im_save restore}bd
/:g
{
1 setgray
0 0 1 1 rF
0 setgray
iw ih polarity[iw 0 0 ih 0 0]setupimageproc
imagemask
}bd
/:h
{
setrgbcolor
0 0 1 1 rF
setrgbcolor
iw ih polarity[iw 0 0 ih 0 0]setupimageproc
imagemask
}bd
/:i
{
setrgbcolor
iw ih polarity[iw 0 0 ih 0 0]setupimageproc
imagemask
}bd
level2  endnoload level2 not startnoload
/$j 9 dict dup
begin
/ImageType 1 def
/MultipleDataSource false def
end
def
/im_save Z
/setupimageproc Z
/polarity Z
/smoothflag Z
/bpc Z
/ih Z
/iw Z
/datatype Z
/:f
{
/im_save save store
/datatype xs
datatype 0 lt datatype 4 gt or{
(error, datatype out of range)print flush stop
}if
/setupimageproc{
{
currentfile
}
{
currentfile 0(
}
{
currentfile/RunLengthDecode filter
}
{
currentfile/ASCII85Decode filter/RunLengthDecode filter
}
{
currentfile/ASCII85Decode filter
}
}datatype get store
{
[1 0]
}{
[0 1]
}ifelse
/polarity xs
/smoothflag xs
:T
scale
pop
/bpc xs
/ih xs
/iw xs
$c setcolorspace
}bd
/:j{im_save restore}bd
/:g
{
1 G
0 0 1 1 rF
0 G
$j dup begin
/Width iw def
/Height ih def
/Decode polarity def
/ImageMatrix[iw 0 0 ih 0 0]def
/DataSource setupimageproc def
/BitsPerComponent 1 def
/Interpolate smoothflag def
end
imagemask
}bd
/:h
{
:F
0 0 1 1 rF
:F
$j dup begin
/Width iw def
/Height ih def
/Decode polarity def
/ImageMatrix[iw 0 0 ih 0 0]def
/DataSource setupimageproc def
/BitsPerComponent 1 def
/Interpolate smoothflag def
end
imagemask
}bd
/:i
{
:F
$j dup begin
/Width iw def
/Height ih def
/Decode polarity def
/ImageMatrix[iw 0 0 ih 0 0]def
/DataSource setupimageproc def
/BitsPerComponent 1 def
/Interpolate smoothflag def
end
imagemask
}bd
level2 not endnoload
/junk Z
/$z Z
userdict/sc_smooth known not
save
systemdict/eexec known
systemdict/cexec known and{
countdictstack mark
false
<1861AEDAE118A9F95F1629C0137F8FE656811DD93DFBEA65E947502E78BA12284B8A58EF0A3
2E272778DAA2ABEC72A84102D591E11D96BA61F57877B895A752D9BEAAC3DFD7D3220E2BDE7
C036467464E0E836748F1DE7AB6216866F130CE7CFCEC8CE050B870C11881EE3E9D70919>
{eexec}stopped{
cleartomark
countdictstack exch sub dup 0 gt{{end}repeat}{pop}ifelse
false
}{
{cleartomark pop true}{cleartomark pop false}ifelse
}ifelse
}{false}ifelse
exch restore and
level2 not and
vmstatus exch sub exch pop 15000 gt and
{
currentfile eexec
}{
/junk save store
/$z 4795 string store
currentfile $z readhexstring pop pop
{
currentfile $z readline not
{
stop
}if
(
{
exit
}if
}bind loop
junk restore
}ifelse
bc89dd93a62e673f17beaf79fc308801f2548cc0804b6e7e7211db7d71dcacee61d4db4b
cc4c192da6ec1c558421396b4eb1944f656db0dda1626374294528747cd1ee8e10b15c5c
60b4182960a4687e44c92cff1b5d29a6b48ab8be9f8115c642241a4901d75a2b2ba55d27
0620b884f37689503d9c3a603e89a1f7de7447e2b23145af7219c13aad065fe60313c4f1
7d1959166b8493e26ee828d6e76ac472747b40432571d2d101dd3e8696849eb59b70c328
a0d1978eea220593cb3024a3cdecb89dcfa2d5ab3990b0bbc1a0c3351bfbd614917e7ede
ef79cb8843325e4a81a80e809250ce8cf7297b5c684b53a56538b373cb085ec7436f82a2
e48b4789de5ac368defd97ca81e1e7a584b7e40bcf852c3d4f6cc387172784295be04ca2
0793987d64efc3cec658553cbe610fa9ebfe74341192cfcc6ecfd0a4843b740cbfd5ba5f
4c076050268792190676f76cacc26be628f8ae1c48419803c2a5108f6b1bec6664b06248
6a083d8092cb3c82b90bded3eed0387490fe971d6614c5d0259a846d43abb22e0dc06aa5
6911c9f53cf5524e138662db3fa2e6cdfe30873d916ed70e4962ed78b32491bee9a20a36
8be439dc6245d5751e6899e2214c07df7a87b66b1aba9a8196db2add3d3c4e3057dc9a9c
dae1cc7884a6f29a5568b1ad64143fd479b8b4a8fbef4db889fe42edaadef0276f79d192
245b64299ad347b03ecf887db96b16fc319a1c7e59718ac7ff7bc7bafd4523bd88fd5ba8
1730817c2f529d3060cb591fe565d778fc989e7e14f7c2a7b85785c53af35fbe738da838
cdf80c9a048f4d7dbd58e2c9f527d2d3b2ad633da1005036298ec8533b94f84e9246289c
f03336f4a7f657afbcbdd7b5e16187273c459f5626a5667db4fbf8b85ef2363ee6d89bd0
f402408cae3065fecf609fa290047e9656d853739f33b06d64a4a57f66f37f7da39fa89d
28962fddb76e8d05683c090664175dda6a1be57698894fd9f0f9d8da411b3f33b3b9c8d4
50a27c37e4a0e503418cd6c02cf60fa7fbb6f6f3e7509a45a54343ba3361168d895a27eb
6923ab65b019b188fe21c804629f2be2a20e6417841b4c8d5d68fb38db71ac7174e68d03
0611ea29f9ca740d0aef611a1f0003e1d136a539ed28d956a2a36bc7ae0ee290bd818175
3537e39be7777de1004964c769ec7e933b17c852cbd2da4cc100d9d3e141106eea9cc29a
b1c910ead768527a02711ac035262c93a4548f67fb501a971e0a86d819547bac7c09f5fb
aec454c9c7276f5e317cc8fd8876e7f90dc128f03052a756cf6db9876fe5a31c9d6a139b
d46eb34272ff7992b86c88420ab07e801a39e91afb3c048b40913dc2d984e0c347b3daea
4e1ce5a15c8df7b65a0b89136f6a48a92b8f096d708a2bea4390f0454dcc3d9cd4f6b24d
8f06faa17694172213c481ceaa1f7fe33f1389142f017874560e1b0272385585fc5681de
4a909566cdcf9af80dfde23e9ad55083bb9422ae57b99bf3d5f081a4cbe0172f236abbb5
06fbbee46987543fc545e81ecd61477fadec55640ce9e41f8cbcd409f0e64c1b83193885
dffe0c4c383de781943a2a73ad1e0cbee0dccdbc3bbbc010f3adc8aa597210cae47003d2
952b8e874e1dee33ef1a78180a3ef6b22e1a66385147f550678510b15ed1a02fc85e736e
818f03b188d6a23f070e5720195a8e4c295d27129a5adedc1a95145d5e758649acebaec5
3d14cbc9feaf3abcecc976c03d00ea640c3b852baad7d8ab28c593cb74cbe2e98ea0b35c
8827eb3def1a79af837fb8468b2042acaf226b1f6d11abab2884f3fe49772325d273f893
82badc7b3b26e43c3f6170eec2c607cbe22225b9469de5509d31bafa1729c416089aeb1b
3b5477a985acdce47557bf29130d6232d2003a4314bf46c9c9383c437f1f2597f361405a
b92f371458f13d8d400d78140df094a559bec9d240993f2d811f0d4eba7678773bb6c765
caf33bc77c0e2e156538b04423e513cb933d9ac8cec0444e0309bdb5c70f02a71f49cc99
7b8d98ecbaad1d53daa2811b07df94b2ec574326d760892fd1886703eed816bf2db96bbe
f2ea914cef918d251a1203d93d549aa438a1309ce179c233111175d9e8bb6064dc2d52db
0976650b1d5d194eab20ed3e500db4410852390e69e56e4ab71e97a87680fa620d8d32d5
b93e40cdec16ed20af734e792ddb938b8a65ccb811f369cc1e9e8819e76d7908e310e5ea
018c05d2db74abc0e8e5da75431324a3ba298820009d6fcc9d5693ec900aab8745112928
ef24a8da713dc8ecebb84394f0058335dafb459971cf20e8c15d40f35643a0c56c95bd86
faefe8e251fb9f79a6db5e481032f812454181a73375ab1efe575d745fe7073adaf60865
e992c7f5e969f4e267b323bb022a7f8b1b480a97ca735212b4d47aff196f37fa177c765a
f03487b323f2ce11314906d04dcb0c3c770ea581fcff70cc1553b4615a6d8dfd69001581
77a9fcc035cb44d630b99cdae09de33c09c9ab8bcc021e72fd88cccf3a9cd578a34a3f96
ae636047fc5f468d22cea0cf509e1588563c7e25d4f0d41b30373dfd1cb69a8e37a51a57
5d1cab7a3c9d04f3d388db04ed78d3e81e7b53dffa94507df197e467565d4169fc996e46
671f62236b00f4e9dc36bd3709400771ed4b335bcdb3c2974c0a00c03717ed980e98dc1b
2473f38bd1e1fe4da0fea5169aa1e06054315579a5898bbe3c9fabb829485562e9d25248
142575f3bfd311dde2297c7883b0b9ed0c3ffe3358880c8c77be01cd6408dc743634006f
d888ffc1e4562ed1166bbdcbe971a269afae52a8a384eaf8880433a053bd072accc62214
04bb18e98b9cebb3c05e56216b56174bd432e41b8c82c82d7114adb16ad941fe87c0657e
03203fc26ed43096e7e262a8f86fc07506b3ffcb5888bcbd3482977716a337a54c0649c1
2579b4322afcb34486a5593fb9d58cb16a0368f593c3c52608bf31c6a3b48a04be605c60
b15ea2670c3e89d293ba065db7d5471d3f56d7ffe3ba48814d5a1f4a81d9865d1f2340cc
3e5ced23ce6827065312bb9ae13e0d3cc53e79aaa764a229be203d1b45963d17e983f72a
f97bac50984a360768f459585745d94f7b77457bb8bc802332aef4e65ae7200d94750aea
b4ede71ade6ee000cf122866f514747784f7e497a6e48b82467515f474d38ec690bac331
d8246eacf2b10d7f1b1b74b1020843b3d934115e23fc1845e5161f5d494badc22f5e68fa
4d1cb3933ac6467e3cc58099adcf4832c156ce1714ee44c6368dee76483e47260dda3af1
97f938530475ede86f1b7ccfeeaeab2518f1c05ba983ada09c069d0f5e7e34c7832ec4c1
9e68e78a51e25540cadf7b64d1eca1141816aba211cea9b8aa4ff9542f26cd6adabe28bc
03c20d2fded392cc160e5ca6f8216611e41b504567927a560537a57e09dd8db552fab25d
59318ad3f7e6b6e10ad49fbb4caf8d5205d494666f5c51a599df65d2c89fe65c92ec733d
78205e987659e02d244684cff18d2b6f1d0d33ccae32339b840246a50cccebc952c7e705
d30fcb5dc94da2eab7ba3c7f9fe536eba4971a8b8e0744a0a094d531ae611042c347fac7
3b770c445121464594a2c5eb9d923f0fe2f33500b403e4b5daf64acd4a4a89d8469be57f
a393a844e082dec5af117b77e7be0eb6474eb47b00503bc68c809f0de147f7d0f1f68798
89b1781d09c0dc12c882dba17e23de1311321fa851a94c51a8801a4bf7c888e4805b1fc8
2fe638aaabb33b13ab0ae024b0b767108a068d562ebd11a91ad8706df9360a914927b752
ddb4d176f72661d12a6349a58a8d688b311ec8e1797b011dbb47760765b88a651b42cace
e60730553040e64711fff10e3285171ed1dae0545b6cfe3503b24f9d135d04d467372b11
ef3c1b9aa7c4272e281df0b6e7cc7e7984b00cbda5bbd81c939213c32176b5e062660e54
8943afe1e39a38cb171d7591c01ac4eea3963572ce008d48727ec8efc0ead1cc4c86d9ce
3d9d5c21f58a41359f2599d4cb15762366c3d3c5b7abe7970a312ea2fb14f5e74f78284a
624fe71f362f4b4d892ac6a87cf8794982964aa0e6153477da1484d35787edb8a2aad90b
3cada91934c118098617e1254d834d86154e6585628663741ae0d7bf339b168d4acec9da
94c12bd5449ecf97a7483e27c47160f838ac076f79b82fc0bd8a3b0011bd0bbb2e6a38ff
a7a63daf463cacb10f56c459bec699ecbcc52be2e5c57bd013c4682b01199cdd89ad2bd1
8584b05e6178c2d6fdc6e19a375ae4a91428d4c0aff2d07e5b6d159351475e57419f2ec8
a343d209e36ddb078027e03d4f179fbf675373c485be1a147ffda00821aba79187471ef9
55f197d6bf0f442244d5b4d1e8392ba11965b6bfa8db30c0e48984883a9b6ec840a9b6fc
93df40c82906dac253e0afd633c336cedac8dd7213e4766761ec3cf3a8aacb1fe1bf2d61
d64c5addaf1d836e199df0ef5e61029f1e6a28c17eba0d8bdf81f358058086ee365b4127
a1bb4b273b84a825fdcc4ebd3abdd7ca243788e1eb48aed96626bd2f2ef8a2e1a082acb6
64b5af60797d9ad021ac616e4d65c1ddbc4a74928b7ab089a10db9312a18d682cccce0b5
53e2f454af0ca21f59b96ee7d71a97db2607dba56daad1bef91c66c44f6fc53d2b4bfebb
876cd478243a9454e8864f65067787b96cb4b3f92a7c90f959ce3324c94590348a8d2709
f4feea7adaf4c56199da203d8f4b9866fe7362d86401b0cfe1913476eb9f7f4d4df0f2f3
8e7ee153c99dda1ccc6eec2cd1b76c52c553ceca04ad525fc17a4a7af197d158c02e6469
cd7e3be4b934ed0e1499153bb98b4363159a60c9ce892af085511cfc31cfa7fa1baf0e22
d138b6a54e93532486b633024d18afd6a610ede93b2a554abc44da890144ff060d1db1e3
e57df15229afe56c49e3c7f68e6ef0221942945f3c6675b498e94fd834bfcd869bba0a90
179d3eb859ea6b749871a62c0e8513ab69b63e35b5bf8060f25442a6670293cee9a86928
997a54929695b3eda64483ecdf8cfb2732729a652e08a3aa82ecb5ac45aad7227f7f7ee9
1d8c222933cfac18561f27939ca29fbf5eebd634555856519cd2f1573281514aff3c6091
5142196c57177ef2086958fbb3fae8f448118c435ae82b7b17a8445e0806b8442e1f9347
c7940af7342dc3d5638958f9a3952768f6c244920e281a163cc87fbdbec99080b8d7e24b
9e35e152aa600a6a4faebb140857e536bb819f4cd9992b9dfaa60ac33a8ff2ed14c791e7
0290e5f6d810488d672224bfeeca6907b41f46489091322b6830a8424fa2a0061151ee61
ef1cfb1a83c4538025605554ed3c734fd8e39bd6da4af6e5a57593da83fd9f511af49b84
5ef7cb34495b54070ea204daff2e266f5a1619928e73de256979f4afdeb42b889064461b
f7263f82a2c233d113de9ba4bc415ed68241bc3862a34050479aa86eabb80cef561ad640
80be73a5da7fbb4905bee2fbbbc7c85a5512c6e747a5eb18942e47c17e5da1127910d8e6
ed7a37d3b846cfda6b0adc8dd594ffe14995b32babcb4d99869090abd57e8ab397e33406
2c7790b673d184738a9e9d578f54ccfcdcc46200d2bb07c5dea46261680dccb4f011428f
289c2049ff81040685aec9ae8c9016c491d02f14f2c4fd79e053147520d59d034c470569
e602c662af4aafc5d2fb0d3af218b26b769ac25fe9999ba3c2eaa8a9ffa93a6f4b743608
f5797772e53cfd2f3c6a45b1a843fcfc6aedc66bc35db923bea051301fa5a959a11eb0b6
f02f7c4e6b004ce548b77e032e36f93cce55f507d5892b32da81a7ce5016326aacd06dd5
0a1b3c615abf06ab6eb1bd643f3c3d7fbc9e18bc11ee29b2113d5c458f20713d3b811da8
883f8b95557ffb1eaf229b89fc98f7623ff9140bcebc6b8677ad7d99373e5757292a3d6a
83a5c053e61224b4dc14e42626637e714c514092e569101ce75f0533605802455f03bf0e
336c0f9788569f3b2b4b79c01628abc3b243decb3c55e1273ff7b83ae61d936bd413580f
3f279437da1fdad63d785ff7bedd815a6d6e4964231adf28640794c33a30686e5f60fe78
c5e87ffd27a84452dc111bbd7c2d0b4e7781c102547f5533ea676a33b8d58918f197a8cd
a1e90be9620e43af9d5b07ee2ae5d702683c9a6973fdaafaec74d62cb29b18272b3f8840
a1a8372ddf4bfee1b7ea037ed41fdc2cb1cb4da3ccb3e121354db0d5bc83c4d4c323083d
75dd32c20607f81fcd221751c5af21775659e804b9941cf6fbe4c8846be20e546f88a588
7bbff8cab4489883fdd2be613313174c015d69eee40eee4242e173aaac87585597feff69
a64f4ef2dbf00dbfe9ad946892816711c7199d4068d8494ba1df614d9bbe1c7b63996e98
12b913966a42a41be0dc85feff232e625f4dd0b01cb7674b232e89d5da143f9d12a9d8a9
aa8b62038721ad9e4a6438cb842d1828458fc25e5b0b6ea795f7a14d2f55f1451c615f4c
211469611d1348952e655e1d93188c0dc6dd719630a42fe5baf8cde0c03b1e0aa76a2cd4
fec86a3042f8a2ea5af50a43af52c1217f6be45b152e47d165b89a54ede432f1eb3ed3ef
428c228dd53c44db15541704b99413b465ff65ca0e9d6be932a18eca713c71ba7dc89238
1e0c2c62cf8567df2f997e58ed2c1ffe17d2fc96db942ce3950e28d1287b85ff7d07adb9
059629dc89b9b0ed2213e45d8a20188cae18f3391887387bf8ba0a12fe21fa0aa521bfa3
718f7abe76388e
0000000000000000000000000000000000000000000000000000000000000000
0000000000000000000000000000000000000000000000000000000000000000
0000000000000000000000000000000000000000000000000000000000000000
0000000000000000000000000000000000000000000000000000000000000000
0000000000000000000000000000000000000000000000000000000000000000
0000000000000000000000000000000000000000000000000000000000000000
0000000000000000000000000000000000000000000000000000000000000000
cleartomark

/@a
{
np :M 0 rl :L 0 exch rl 0 rl :L fill
}bd
/@b
{
np :M 0 rl 0 exch rl :L 0 rl 0 exch rl fill
}bd
/arct where
{
pop
}{
/arct
{
arcto pop pop pop pop
}bd
}ifelse
/x1 Z
/x2 Z
/y1 Z
/y2 Z
/rad Z
/@q
{
/rad xs
/y2 xs
/x2 xs
/y1 xs
/x1 xs
np
x2 x1 add 2 div y1 :M
x2 y1 x2 y2 rad arct
x2 y2 x1 y2 rad arct
x1 y2 x1 y1 rad arct
x1 y1 x2 y1 rad arct
fill
}bd
/@s
{
/rad xs
/y2 xs
/x2 xs
/y1 xs
/x1 xs
np
x2 x1 add 2 div y1 :M
x2 y1 x2 y2 rad arct
x2 y2 x1 y2 rad arct
x1 y2 x1 y1 rad arct
x1 y1 x2 y1 rad arct
:K
stroke
}bd
/@i
{
np 0 360 arc fill
}bd
/@j
{
gS
np
:T
scale
0 0 .5 0 360 arc
fill
gR
}bd
/@e
{
np
0 360 arc
:K
stroke
}bd
/@f
{
np
$m currentmatrix
pop
:T
scale
0 0 .5 0 360 arc
:K
$m setmatrix
stroke
}bd
/@k
{
gS
np
:T
0 0 :M
0 0 5 2 roll
arc fill
gR
}bd
/@l
{
gS
np
:T
0 0 :M
scale
0 0 .5 5 -2 roll arc
fill
gR
}bd
/@m
{
np
arc
stroke
}bd
/@n
{
np
$m currentmatrix
pop
:T
scale
0 0 .5 5 -2 roll arc
$m setmatrix
stroke
}bd
/S/show ld
/A{
0.0 exch ashow
}bd
/R{
0.0 exch 32 exch widthshow
}bd
/W{
0.0 3 1 roll widthshow
}bd
/J{
0.0 32 4 2 roll 0.0 exch awidthshow
}bd
/V{
0.0 4 1 roll 0.0 exch awidthshow
}bd
/fcflg true def
/fc{
fcflg{
vmstatus exch sub 50000 lt{
(
}if pop
}if
}bd
/$f[1 0 0 -1 0 0]def
/:ff{$f :mf}bd
/MacEncoding StandardEncoding 256 array copy def
MacEncoding 39/quotesingle put
MacEncoding 96/grave put
/Adieresis/Aring/Ccedilla/Eacute/Ntilde/Odieresis/Udieresis/aacute
/agrave/acircumflex/adieresis/atilde/aring/ccedilla/eacute/egrave
/ecircumflex/edieresis/iacute/igrave/icircumflex/idieresis/ntilde/oacute
/ograve/ocircumflex/odieresis/otilde/uacute/ugrave/ucircumflex/udieresis
/dagger/degree/cent/sterling/section/bullet/paragraph/germandbls
/registered/copyright/trademark/acute/dieresis/notequal/AE/Oslash
/infinity/plusminus/lessequal/greaterequal/yen/mu/partialdiff/summation
/product/pi/integral/ordfeminine/ordmasculine/Omega/ae/oslash
/questiondown/exclamdown/logicalnot/radical/florin/approxequal/Delta/guillemotleft
/guillemotright/ellipsis/space/Agrave/Atilde/Otilde/OE/oe
/endash/emdash/quotedblleft/quotedblright/quoteleft/quoteright/divide/lozenge
/ydieresis/Ydieresis/fraction/currency/guilsinglleft/guilsinglright/fi/fl
/daggerdbl/periodcentered/quotesinglbase/quotedblbase/perthousand
/Acircumflex/Ecircumflex/Aacute/Edieresis/Egrave/Iacute/Icircumflex/Idieresis/Igrave
/Oacute/Ocircumflex/apple/Ograve/Uacute/Ucircumflex/Ugrave/dotlessi/circumflex/tilde
/macron/breve/dotaccent/ring/cedilla/hungarumlaut/ogonek/caron
MacEncoding 128 128 getinterval astore pop
level2 startnoload
/copyfontdict
{
findfont dup length dict
begin
{
1 index/FID ne{def}{pop pop}ifelse
}forall
}bd
level2 endnoload level2 not startnoload
/copyfontdict
{
findfont dup length dict
copy
begin
}bd
level2 not endnoload
md/fontname known not{
/fontname/customfont def
}if
/Encoding Z
/:mre
{
copyfontdict
/Encoding MacEncoding def
fontname currentdict
end
definefont :ff def
}bd
/:bsr
{
copyfontdict
/Encoding Encoding 256 array copy def
Encoding dup
}bd
/pd{put dup}bd
/:esr
{
pop pop
fontname currentdict
end
definefont :ff def
}bd
/scf
{
scalefont def
}bd
/scf-non
{
$m scale :mf setfont
}bd
/ps Z
/fz{/ps xs}bd
/sf/setfont ld
/cF/currentfont ld
/mbf
{
/makeblendedfont where
{
pop
makeblendedfont
/ABlend exch definefont
}{
pop
}ifelse
def
}def
/wi
version(23.0)eq
{
{
gS 0 0 0 0 rC stringwidth gR
}bind
}{
/stringwidth load
}ifelse
def
/$o 1. def
/gl{$o G}bd
/ms{:M S}bd
/condensedmtx[.82 0 0 1 0 0]def
/:mc
{
condensedmtx :mf def
}bd
/extendedmtx[1.18 0 0 1 0 0]def
/:me
{
extendedmtx :mf def
}bd
/basefont Z
/basefonto Z
/dxa Z
/dxb Z
/dxc Z
/dxd Z
/dsdx2 Z
/bfproc Z
/:fbase
{
dup/FontType get 0 eq{
dup length dict begin
dup{1 index/FID ne 2 index/UniqueID ne and{def}{pop pop}ifelse}forall
/FDepVector exch/FDepVector get[exch/:fbase load forall]def
}/bfproc load ifelse
/customfont currentdict end definefont
}bd
/:mo
{
/bfproc{
dup dup length 2 add dict
begin
{
1 index/FID ne 2 index/UniqueID ne and{def}{pop pop}ifelse
}forall
/PaintType 2 def
/StrokeWidth .012 0 FontMatrix idtransform pop def
/customfont currentdict
end
definefont
8 dict begin
/basefonto xdf
/basefont xdf
/FontType 3 def
/FontMatrix[1 0 0 1 0 0]def
/FontBBox[0 0 1 1]def
/Encoding StandardEncoding def
/BuildChar
{
exch begin
basefont setfont
( )dup 0 4 -1 roll put
dup wi
setcharwidth
0 0 :M
gS
gl
dup show
gR
basefonto setfont
show
end
}def
}store :fbase
}bd
/:mso
{
/bfproc{
7 dict begin
/basefont xdf
/FontType 3 def
/FontMatrix[1 0 0 1 0 0]def
/FontBBox[0 0 1 1]def
/Encoding StandardEncoding def
/BuildChar
{
exch begin
sD begin
/dxa 1 ps div def
basefont setfont
( )dup 0 4 -1 roll put
dup wi
1 index 0 ne
{
exch dxa add exch
}if
setcharwidth
dup 0 0 ms
dup dxa 0 ms
dup dxa dxa ms
dup 0 dxa ms
gl
dxa 2. div dup ms
end
end
}def
}store :fbase
}bd
/:ms
{
/bfproc{
dup dup length 2 add dict
begin
{
1 index/FID ne 2 index/UniqueID ne and{def}{pop pop}ifelse
}forall
/PaintType 2 def
/StrokeWidth .012 0 FontMatrix idtransform pop def
/customfont currentdict
end
definefont
8 dict begin
/basefonto xdf
/basefont xdf
/FontType 3 def
/FontMatrix[1 0 0 1 0 0]def
/FontBBox[0 0 1 1]def
/Encoding StandardEncoding def
/BuildChar
{
exch begin
sD begin
/dxb .05 def
basefont setfont
( )dup 0 4 -1 roll put
dup wi
exch dup 0 ne
{
dxb add
}if
exch setcharwidth
dup dxb .01 add 0 ms
0 dxb :T
gS
gl
dup 0 0 ms
gR
basefonto setfont
0 0 ms
end
end
}def
}store :fbase
}bd
/:mss
{
/bfproc{
7 dict begin
/basefont xdf
/FontType 3 def
/FontMatrix[1 0 0 1 0 0]def
/FontBBox[0 0 1 1]def
/Encoding StandardEncoding def
/BuildChar
{
exch begin
sD begin
/dxc 1 ps div def
/dsdx2 .05 dxc 2 div add def
basefont setfont
( )dup 0 4 -1 roll put
dup wi
exch dup 0 ne
{
dsdx2 add
}if
exch setcharwidth
dup dsdx2 .01 add 0 ms
0 .05 dxc 2 div sub :T
dup 0 0 ms
dup dxc 0 ms
dup dxc dxc ms
dup 0 dxc ms
gl
dxc 2 div dup ms
end
end
}def
}store :fbase
}bd
/:msb
{
/bfproc{
7 dict begin
/basefont xdf
/FontType 3 def
/FontMatrix[1 0 0 1 0 0]def
/FontBBox[0 0 1 1]def
/Encoding StandardEncoding def
/BuildChar
{
exch begin
sD begin
/dxd .03 def
basefont setfont
( )dup 0 4 -1 roll put
dup wi
1 index 0 ne
{
exch dxd add exch
}if
setcharwidth
dup 0 0 ms
dup dxd 0 ms
dup dxd dxd ms
0 dxd ms
end
end
}def
}store :fbase
}bd
/italicmtx[1 0 -.212557 1 0 0]def
/:mi
{
italicmtx :mf def
}bd
/:v
{
[exch dup/FontMatrix get exch
dup/FontInfo known
{
/FontInfo get
dup/UnderlinePosition known
{
dup/UnderlinePosition get
2 index 0
3 1 roll
transform
exch pop
}{
.1
}ifelse
3 1 roll
dup/UnderlineThickness known
{
/UnderlineThickness get
exch 0 3 1 roll
transform
exch pop
abs
}{
pop pop .067
}ifelse
}{
pop pop .1 .067
}ifelse
]
}bd
/$t Z
/$p Z
/$s Z
/:p
{
aload pop
2 index mul/$t xs
1 index mul/$p xs
.012 mul/$s xs
}bd
/:m
{gS
0 $p rm
$t lw
0 rl stroke
gR
}bd
/:n
{
gS
0 $p rm
$t lw
0 rl
gS
gl
stroke
gR
strokepath
$s lw
/setstrokeadjust where{pop
currentstrokeadjust true setstrokeadjust stroke setstrokeadjust
}{
stroke
}ifelse
gR
}bd
/:o
{gS
0 $p rm
$t 2 div dup rm
$t lw
dup 0 rl
stroke
gR
:n
}bd
/currentpacking where {pop sc_oldpacking setpacking}if
end		
md begin
countdictstack
[
{
level2 {1 dict dup /ManualFeed false put setpagedevice}{statusdict begin
/manualfeed false store end} ifelse
}featurecleanup
countdictstack
[
{
}featurecleanup
countdictstack
[
{
level2 {
		2 dict dup /PageSize [612 792] put dup /ImagingBBox null put setpagedevice
	}{
		/lettersmall where {pop lettersmall} {letterR} ifelse
	} ifelse
}featurecleanup
(Norman-B32613A-P049064)setjob
/pT[1 0 0 -1 29.999 761.009]def/mT[.24 0 0 -.24 29.999 761.009]def
/sD 16 dict def
300 level2{1 dict dup/WaitTimeout 4 -1 roll put
setuserparams}{statusdict/waittimeout 3 -1 roll put}ifelse
/f0_1/Helvetica :mre
/f1_1 f0_1 1.04 scf
/f1_12 f1_1 12 scf
/f3_1/Helvetica-Bold :mre
/f4_1 f3_1 1.04 scf
/f4_14 f4_1 14 scf
/f5_1/Symbol :bsr
240/apple pd
:esr /f5_12 f5_1 12 scf
/Courier findfont[10 0 0 -10 0 0]:mf setfont
initializepage
(Norman-B32613A-P049064; page: 1 of 1)setjob
gS 0 0 2300 3042 rC
1 G
391 1495 1578 1171 rF
391 308 1578 1172 rF
0 G
704 1728 1049 780 rC
-1 -1 706 2508 1 1 725 2495 @b
-1 -1 726 2496 1 1 746 2482 @b
-1 -1 747 2483 1 1 767 2469 @b
-1 -1 768 2470 1 1 788 2456 @b
-1 -1 789 2457 1 1 809 2442 @b
-1 -1 810 2443 1 1 830 2430 @b
-1 -1 831 2431 1 1 851 2417 @b
-1 -1 852 2418 1 1 872 2404 @b
-1 -1 873 2405 1 1 893 2391 @b
-1 -1 894 2392 1 1 914 2378 @b
-1 -1 915 2379 1 1 935 2365 @b
-1 -1 936 2366 1 1 956 2352 @b
-1 -1 957 2353 1 1 977 2339 @b
-1 -1 978 2340 1 1 998 2326 @b
-1 -1 999 2327 1 1 1019 2313 @b
-1 -1 1020 2314 1 1 1040 2300 @b
-1 -1 1041 2301 1 1 1060 2287 @b
-1 -1 1061 2288 1 1 1082 2274 @b
-1 -1 1083 2275 1 1 1102 2261 @b
-1 -1 1103 2262 1 1 1124 2248 @b
-1 -1 1125 2249 1 1 1144 2235 @b
-1 -1 1145 2236 1 1 1165 2222 @b
-1 -1 1166 2223 1 1 1186 2209 @b
-1 -1 1187 2210 1 1 1207 2196 @b
-1 -1 1208 2197 1 1 1228 2183 @b
-1 -1 1229 2184 1 1 1249 2170 @b
-1 -1 1250 2171 1 1 1270 2157 @b
-1 -1 1271 2158 1 1 1291 2144 @b
-1 -1 1292 2145 1 1 1312 2131 @b
-1 -1 1313 2132 1 1 1333 2118 @b
-1 -1 1334 2119 1 1 1354 2105 @b
-1 -1 1355 2106 1 1 1375 2092 @b
-1 -1 1376 2093 1 1 1396 2079 @b
-1 -1 1397 2080 1 1 1417 2066 @b
-1 -1 1418 2067 1 1 1438 2053 @b
-1 -1 1439 2054 1 1 1459 2040 @b
-1 -1 1460 2041 1 1 1480 2027 @b
-1 -1 1481 2028 1 1 1501 2014 @b
-1 -1 1502 2015 1 1 1522 2001 @b
-1 -1 1523 2002 1 1 1543 1988 @b
-1 -1 1544 1989 1 1 1564 1975 @b
-1 -1 1565 1976 1 1 1584 1962 @b
-1 -1 1585 1963 1 1 1606 1949 @b
-1 -1 1607 1950 1 1 1626 1936 @b
-1 -1 1627 1937 1 1 1648 1923 @b
-1 -1 1649 1924 1 1 1668 1910 @b
-1 -1 1669 1911 1 1 1689 1897 @b
-1 -1 1690 1898 1 1 1710 1884 @b
-1 -1 1711 1885 1 1 1731 1871 @b
-1 -1 1732 1872 1 1 1752 1858 @b
-1 -1 705 2455 1 1 725 2449 @b
-1 -1 726 2450 1 1 746 2436 @b
-1 -1 747 2437 1 1 767 2428 @b
-1 -1 768 2429 1 1 788 2417 @b
-1 -1 789 2418 1 1 809 2407 @b
-1 -1 810 2408 1 1 830 2396 @b
-1 -1 831 2397 1 1 851 2386 @b
-1 -1 852 2387 1 1 872 2376 @b
-1 -1 873 2377 1 1 893 2365 @b
-1 -1 894 2366 1 1 914 2354 @b
-1 -1 915 2355 1 1 935 2344 @b
-1 -1 936 2345 1 1 956 2333 @b
-1 -1 957 2334 1 1 977 2323 @b
-1 -1 978 2324 1 1 998 2312 @b
-1 -1 999 2313 1 1 1019 2301 @b
-1 -1 1020 2302 1 1 1040 2291 @b
-1 -1 1041 2292 1 1 1060 2280 @b
-1 -1 1061 2281 1 1 1082 2269 @b
-1 -1 1083 2270 1 1 1102 2259 @b
-1 -1 1103 2260 1 1 1124 2248 @b
-1 -1 1125 2249 1 1 1144 2237 @b
-1 -1 1145 2238 1 1 1165 2226 @b
-1 -1 1166 2227 1 1 1186 2215 @b
-1 -1 1187 2216 1 1 1207 2205 @b
-1 -1 1208 2206 1 1 1228 2194 @b
-1 -1 1229 2195 1 1 1249 2183 @b
-1 -1 1250 2184 1 1 1270 2172 @b
-1 -1 1271 2173 1 1 1291 2161 @b
-1 -1 1292 2162 1 1 1312 2150 @b
-1 -1 1313 2151 1 1 1333 2139 @b
-1 -1 1334 2140 1 1 1354 2129 @b
-1 -1 1355 2130 1 1 1375 2117 @b
-1 -1 1376 2118 1 1 1396 2107 @b
-1 -1 1397 2108 1 1 1417 2096 @b
-1 -1 1418 2097 1 1 1438 2085 @b
-1 -1 1439 2086 1 1 1459 2074 @b
-1 -1 1460 2075 1 1 1480 2063 @b
-1 -1 1481 2064 1 1 1501 2052 @b
-1 -1 1502 2053 1 1 1522 2042 @b
-1 -1 1523 2043 1 1 1543 2031 @b
-1 -1 1544 2032 1 1 1564 2020 @b
-1 -1 1565 2021 1 1 1584 2009 @b
-1 -1 1585 2010 1 1 1606 1998 @b
-1 -1 1607 1999 1 1 1626 1987 @b
-1 -1 1627 1988 1 1 1648 1977 @b
-1 -1 1649 1978 1 1 1668 1966 @b
-1 -1 1669 1967 1 1 1689 1955 @b
-1 -1 1690 1956 1 1 1710 1945 @b
-1 -1 1711 1946 1 1 1731 1935 @b
-1 -1 1732 1936 1 1 1752 1924 @b
-1 -1 705 2385 1 1 725 2381 @b
-1 -1 726 2382 1 1 746 2375 @b
-1 -1 747 2376 1 1 767 2369 @b
-1 -1 768 2370 1 1 788 2361 @b
-1 -1 789 2362 1 1 809 2354 @b
-1 -1 810 2355 1 1 830 2347 @b
-1 -1 831 2348 1 1 851 2340 @b
-1 -1 852 2341 1 1 872 2332 @b
-1 -1 873 2333 1 1 893 2325 @b
-1 -1 894 2326 1 1 914 2318 @b
-1 -1 915 2319 1 1 935 2310 @b
-1 -1 936 2311 1 1 956 2302 @b
-1 -1 957 2303 1 1 977 2295 @b
-1 -1 978 2296 1 1 998 2287 @b
-1 -1 999 2288 1 1 1019 2280 @b
-1 -1 1020 2281 1 1 1040 2272 @b
-1 -1 1041 2273 1 1 1060 2264 @b
-1 -1 1061 2265 1 1 1082 2256 @b
-1 -1 1083 2257 1 1 1102 2248 @b
-1 -1 1103 2249 1 1 1124 2240 @b
-1 -1 1125 2241 1 1 1144 2232 @b
-1 -1 1145 2233 1 1 1165 2224 @b
-1 -1 1166 2225 1 1 1186 2216 @b
-1 -1 1187 2217 1 1 1207 2208 @b
-1 -1 1208 2209 1 1 1228 2200 @b
-1 -1 1229 2201 1 1 1249 2192 @b
-1 -1 1250 2193 1 1 1270 2184 @b
-1 -1 1271 2185 1 1 1291 2176 @b
-1 -1 1292 2177 1 1 1312 2167 @b
-1 -1 1313 2168 1 1 1333 2159 @b
-1 -1 1334 2160 1 1 1354 2151 @b
-1 -1 1355 2152 1 1 1375 2143 @b
-1 -1 1376 2144 1 1 1396 2135 @b
-1 -1 1397 2136 1 1 1417 2126 @b
-1 -1 1418 2127 1 1 1438 2119 @b
-1 -1 1439 2120 1 1 1459 2110 @b
-1 -1 1460 2111 1 1 1480 2102 @b
-1 -1 1481 2103 1 1 1501 2094 @b
-1 -1 1502 2095 1 1 1522 2085 @b
-1 -1 1523 2086 1 1 1543 2077 @b
-1 -1 1544 2078 1 1 1564 2069 @b
-1 -1 1565 2070 1 1 1584 2061 @b
-1 -1 1585 2062 1 1 1606 2053 @b
-1 -1 1607 2054 1 1 1626 2045 @b
-1 -1 1627 2046 1 1 1648 2037 @b
-1 -1 1649 2038 1 1 1668 2029 @b
-1 -1 1669 2030 1 1 1689 2021 @b
-1 -1 1690 2022 1 1 1710 2013 @b
-1 -1 1711 2014 1 1 1731 2005 @b
-1 -1 1732 2006 1 1 1752 1997 @b
-1 -1 705 2326 1 1 725 2319 @b
-1 -1 726 2320 1 1 746 2317 @b
-1 -1 747 2318 1 1 767 2313 @b
-1 -1 768 2314 1 1 788 2310 @b
-1 -1 789 2311 1 1 809 2307 @b
-1 -1 810 2308 1 1 830 2303 @b
-1 -1 831 2304 1 1 851 2299 @b
-1 -1 852 2300 1 1 872 2295 @b
-1 -1 873 2296 1 1 893 2291 @b
-1 -1 894 2292 1 1 914 2287 @b
-1 -1 915 2288 1 1 935 2282 @b
-1 -1 936 2283 1 1 956 2278 @b
-1 -1 957 2279 1 1 977 2273 @b
-1 -1 978 2274 1 1 998 2268 @b
-1 -1 999 2269 1 1 1019 2263 @b
-1 -1 1020 2264 1 1 1040 2259 @b
-1 -1 1041 2260 1 1 1060 2254 @b
-1 -1 1061 2255 1 1 1082 2249 @b
-1 -1 1083 2250 1 1 1102 2244 @b
-1 -1 1103 2245 1 1 1124 2239 @b
-1 -1 1125 2240 1 1 1144 2234 @b
-1 -1 1145 2235 1 1 1165 2229 @b
-1 -1 1166 2230 1 1 1186 2224 @b
-1 -1 1187 2225 1 1 1207 2218 @b
-1 -1 1208 2219 1 1 1228 2213 @b
-1 -1 1229 2214 1 1 1249 2208 @b
-1 -1 1250 2209 1 1 1270 2203 @b
-1 -1 1271 2204 1 1 1291 2198 @b
-1 -1 1292 2199 1 1 1312 2193 @b
-1 -1 1313 2194 1 1 1333 2187 @b
-1 -1 1334 2188 1 1 1354 2182 @b
-1 -1 1355 2183 1 1 1375 2176 @b
-1 -1 1376 2177 1 1 1396 2171 @b
-1 -1 1397 2172 1 1 1417 2166 @b
-1 -1 1418 2167 1 1 1438 2160 @b
-1 -1 1439 2161 1 1 1459 2155 @b
-1 -1 1460 2156 1 1 1480 2150 @b
-1 -1 1481 2151 1 1 1501 2144 @b
-1 -1 1502 2145 1 1 1522 2139 @b
-1 -1 1523 2140 1 1 1543 2134 @b
-1 -1 1544 2135 1 1 1564 2128 @b
-1 -1 1565 2129 1 1 1584 2122 @b
-1 -1 1585 2123 1 1 1606 2117 @b
-1 -1 1607 2118 1 1 1626 2112 @b
-1 -1 1627 2113 1 1 1648 2106 @b
-1 -1 1649 2107 1 1 1668 2101 @b
-1 -1 1669 2102 1 1 1689 2096 @b
-1 -1 1690 2097 1 1 1710 2091 @b
-1 -1 1711 2092 1 1 1731 2085 @b
-1 -1 1732 2086 1 1 1752 2080 @b
-1 -1 705 2271 1 1 725 2267 @b
-1 -1 726 2268 1 1 746 2266 @b
-1 -1 747 2267 1 1 767 2265 @b
-1 -1 768 2266 1 1 788 2264 @b
-1 -1 789 2265 1 1 809 2263 @b
-1 -1 810 2264 1 1 830 2262 @b
-1 -1 831 2263 1 1 851 2261 @b
-1 -1 852 2262 1 1 872 2259 @b
-1 -1 873 2260 1 1 893 2258 @b
-1 -1 894 2259 1 1 914 2256 @b
-1 -1 915 2257 1 1 935 2254 @b
-1 -1 936 2255 1 1 956 2251 @b
-1 -1 957 2252 1 1 977 2249 @b
-1 -1 978 2250 1 1 998 2247 @b
-1 -1 999 2248 1 1 1019 2245 @b
-1 -1 1020 2246 1 1 1040 2243 @b
-1 -1 1041 2244 1 1 1060 2240 @b
-1 -1 1061 2241 1 1 1082 2237 @b
-1 -1 1083 2238 1 1 1102 2235 @b
-1 -1 1103 2236 1 1 1124 2233 @b
-1 -1 1125 2234 1 1 1144 2230 @b
-1 -1 1145 2231 1 1 1165 2228 @b
-1 -1 1166 2229 1 1 1186 2225 @b
-1 -1 1187 2226 1 1 1207 2222 @b
-1 -1 1208 2223 1 1 1228 2219 @b
-1 -1 1229 2220 1 1 1249 2217 @b
-1 -1 1250 2218 1 1 1270 2214 @b
-1 -1 1271 2215 1 1 1291 2211 @b
-1 -1 1292 2212 1 1 1312 2209 @b
-1 -1 1313 2210 1 1 1333 2205 @b
-1 -1 1334 2206 1 1 1354 2203 @b
-1 -1 1355 2204 1 1 1375 2200 @b
-1 -1 1376 2201 1 1 1396 2197 @b
-1 -1 1397 2198 1 1 1417 2195 @b
-1 -1 1418 2196 1 1 1438 2191 @b
-1 -1 1439 2192 1 1 1459 2189 @b
-1 -1 1460 2190 1 1 1480 2186 @b
-1 -1 1481 2187 1 1 1501 2183 @b
-1 -1 1502 2184 1 1 1522 2180 @b
-1 -1 1523 2181 1 1 1543 2177 @b
-1 -1 1544 2178 1 1 1564 2174 @b
-1 -1 1565 2175 1 1 1584 2171 @b
-1 -1 1585 2172 1 1 1606 2168 @b
-1 -1 1607 2169 1 1 1626 2165 @b
-1 -1 1627 2166 1 1 1648 2162 @b
-1 -1 1649 2163 1 1 1668 2159 @b
-1 -1 1669 2160 1 1 1689 2156 @b
-1 -1 1690 2157 1 1 1710 2153 @b
-1 -1 1711 2154 1 1 1731 2150 @b
-1 -1 1732 2151 1 1 1752 2147 @b
704 2208 -1 1 726 2207 1 704 2207 @a
725 2208 -1 1 747 2207 1 725 2207 @a
746 2208 -1 1 768 2207 1 746 2207 @a
-1 -1 768 2208 1 1 788 2206 @b
788 2207 -1 1 810 2206 1 788 2206 @a
-1 -1 810 2207 1 1 830 2205 @b
830 2206 -1 1 852 2205 1 830 2205 @a
851 2206 -1 1 873 2205 1 851 2205 @a
872 2206 -1 1 894 2205 1 872 2205 @a
-1 -1 894 2206 1 1 914 2204 @b
914 2205 -1 1 936 2204 1 914 2204 @a
-1 -1 936 2205 1 1 956 2203 @b
956 2204 -1 1 978 2203 1 956 2203 @a
977 2204 -1 1 999 2203 1 977 2203 @a
998 2204 -1 1 1020 2203 1 998 2203 @a
-1 -1 1020 2204 1 1 1040 2202 @b
1040 2203 -1 1 1061 2202 1 1040 2202 @a
-1 -1 1061 2203 1 1 1082 2201 @b
-1 -1 1083 2202 1 1 1102 2200 @b
1102 2201 -1 1 1125 2200 1 1102 2200 @a
-1 -1 1125 2201 1 1 1144 2199 @b
-1 -1 1145 2200 1 1 1165 2198 @b
1165 2199 -1 1 1187 2198 1 1165 2198 @a
-1 -1 1187 2199 1 1 1207 2197 @b
1207 2198 -1 1 1229 2197 1 1207 2197 @a
-1 -1 1229 2198 1 1 1249 2196 @b
-1 -1 1250 2197 1 1 1270 2195 @b
1270 2196 -1 1 1292 2195 1 1270 2195 @a
-1 -1 1292 2196 1 1 1312 2194 @b
-1 -1 1313 2195 1 1 1333 2193 @b
1333 2194 -1 1 1355 2193 1 1333 2193 @a
-1 -1 1355 2194 1 1 1375 2192 @b
-1 -1 1376 2193 1 1 1396 2191 @b
-1 -1 1397 2192 1 1 1417 2190 @b
-1 -1 1418 2191 1 1 1438 2189 @b
1438 2190 -1 1 1460 2189 1 1438 2189 @a
-1 -1 1460 2190 1 1 1480 2188 @b
-1 -1 1481 2189 1 1 1501 2187 @b
-1 -1 1502 2188 1 1 1522 2186 @b
-1 -1 1523 2187 1 1 1543 2185 @b
1543 2186 -1 1 1565 2185 1 1543 2185 @a
-1 -1 1565 2186 1 1 1584 2184 @b
-1 -1 1585 2185 1 1 1606 2183 @b
-1 -1 1607 2184 1 1 1626 2182 @b
1626 2183 -1 1 1649 2182 1 1626 2182 @a
-1 -1 1649 2183 1 1 1668 2180 @b
1668 2181 -1 1 1690 2180 1 1668 2180 @a
-1 -1 1690 2181 1 1 1710 2179 @b
-1 -1 1711 2180 1 1 1731 2178 @b
-1 -1 1732 2179 1 1 1752 2177 @b
gR
gS 390 1494 1572 1170 rC
704.5 2462.5 14.5 @i
704.5 2462.5 14 @e
704.5 2371.5 14.5 @i
704.5 2371.5 14 @e
704.5 2306.5 14.5 @i
704.5 2306.5 14 @e
704.5 2261.5 14.5 @i
704.5 2261.5 14 @e
704.5 2196.5 14.5 @i
704.5 2196.5 14 @e
704.5 2137.5 14.5 @i
704.5 2137.5 14 @e
903.5 2417.5 14.5 @i
903.5 2417.5 14 @e
903.5 2293.5 14.5 @i
903.5 2293.5 14 @e
903.5 2261.5 14.5 @i
903.5 2261.5 14 @e
903.5 2196.5 14.5 @i
903.5 2196.5 14 @e
1448.5 1949.5 14.5 @i
1448.5 1949.5 14 @e
30 29 1606 2072.5 @j
29 28 1606 2072.5 @f
1626.5 1981.5 14.5 @i
1626.5 1981.5 14 @e
1668.5 2046.5 14.5 @i
1668.5 2046.5 14 @e
1721.5 1981.5 14.5 @i
1721.5 1981.5 14 @e
1721.5 1858.5 14.5 @i
1721.5 1858.5 14 @e
1752.5 1923.5 14.5 @i
1752.5 1923.5 14 @e
1752.5 1825.5 14.5 @i
1752.5 1825.5 14 @e
0 0 1 :F
679 2508 -1 1 705 2507 1 679 2507 @a
679 2379 -1 1 705 2378 1 679 2378 @a
679 2249 -1 1 705 2248 1 679 2248 @a
679 2119 -1 1 705 2118 1 679 2118 @a
679 1989 -1 1 705 1988 1 679 1988 @a
679 1859 -1 1 705 1858 1 679 1858 @a
679 1729 -1 1 705 1728 1 679 1728 @a
1752 2508 -1 1 1778 2507 1 1752 2507 @a
1752 2379 -1 1 1778 2378 1 1752 2378 @a
1752 2249 -1 1 1778 2248 1 1752 2248 @a
1752 2119 -1 1 1778 2118 1 1752 2118 @a
1752 1989 -1 1 1778 1988 1 1752 1988 @a
1752 1859 -1 1 1778 1858 1 1752 1858 @a
1752 1729 -1 1 1778 1728 1 1752 1728 @a
-1 -1 705 2533 1 1 704 2507 @b
-1 -1 915 2533 1 1 914 2507 @b
-1 -1 1125 2533 1 1 1124 2507 @b
-1 -1 1334 2533 1 1 1333 2507 @b
-1 -1 1544 2533 1 1 1543 2507 @b
-1 -1 1753 2533 1 1 1752 2507 @b
-1 -1 705 1729 1 1 704 1703 @b
-1 -1 915 1729 1 1 914 1703 @b
-1 -1 1125 1729 1 1 1124 1703 @b
-1 -1 1334 1729 1 1 1333 1703 @b
-1 -1 1544 1729 1 1 1543 1703 @b
-1 -1 1753 1729 1 1 1752 1703 @b
-1 -1 705 2508 1 1 704 1728 @b
704 2508 -1 1 1753 2507 1 704 2507 @a
-1 -1 1753 2508 1 1 1752 1728 @b
704 1729 -1 1 1753 1728 1 704 1728 @a
1 G
-8 -8 402 1506 8 8 394 1498 @b
gS
4.17 4.163 scale
151.815 607.2 :T
-151.815 -607.2 :T
151.815 607.2 :M
0 G
f1_12 sf
(0)S
141.022 576.215 :M
.829(0.2)A
141.022 544.27 :M
.829(0.4)A
141.022 513.285 :M
.829(0.6)A
141.022 482.301 :M
.829(0.8)A
151.815 451.316 :M
(1)S
141.022 420.332 :M
.829(1.2)A
165.006 623.292 :M
(0)S
gR
gR
1 G
8 lw
gS 0 0 2300 3042 rC
gS
4.17 4.163 scale
118.958 520.251 :T
270 rotate
-118.958 -520.251 :T
118.958 520.251 :M
0 G
f1_12 sf
-.791(Gap)A
gR
390 1494 1572 1170 rC
gS
4.17 4.163 scale
210.814 623.292 :T
-210.814 -623.292 :T
210.814 623.292 :M
0 G
f1_12 sf
.829(0.2)A
260.939 623.292 :M
.829(0.4)A
310.824 623.292 :M
.829(0.6)A
360.95 623.292 :M
.829(0.8)A
416.831 623.292 :M
(1)S
272.931 637.223 :M
.873(cos\(2)A
306.987 637.223 :M
f5_12 sf
.358(f)A
f1_12 sf
(\))S
265.976 406.161 :M
f4_14 sf
2.091(s-wave)A
427.863 450.115 :M
f1_12 sf
.829(0.0)A
428.823 468.13 :M
.829(0.1)A
428.823 486.144 :M
.829(0.3)A
428.823 505.119 :M
.829(0.6)A
428.823 520.251 :M
.829(1.0)A
428.823 533.221 :M
.829(2.0)A
gR
gR
0 G
gS 704 543 1049 780 rC
-1 -1 706 1323 1 1 725 1310 @b
-1 -1 726 1311 1 1 746 1297 @b
-1 -1 747 1298 1 1 767 1284 @b
-1 -1 768 1285 1 1 788 1271 @b
-1 -1 789 1272 1 1 809 1257 @b
-1 -1 810 1258 1 1 830 1245 @b
-1 -1 831 1246 1 1 851 1232 @b
-1 -1 852 1233 1 1 872 1219 @b
-1 -1 873 1220 1 1 893 1206 @b
-1 -1 894 1207 1 1 914 1193 @b
-1 -1 915 1194 1 1 935 1180 @b
-1 -1 936 1181 1 1 956 1167 @b
-1 -1 957 1168 1 1 977 1154 @b
-1 -1 978 1155 1 1 998 1141 @b
-1 -1 999 1142 1 1 1019 1128 @b
-1 -1 1020 1129 1 1 1040 1115 @b
-1 -1 1041 1116 1 1 1060 1102 @b
-1 -1 1061 1103 1 1 1082 1089 @b
-1 -1 1083 1090 1 1 1102 1076 @b
-1 -1 1103 1077 1 1 1124 1063 @b
-1 -1 1125 1064 1 1 1144 1050 @b
-1 -1 1145 1051 1 1 1165 1037 @b
-1 -1 1166 1038 1 1 1186 1024 @b
-1 -1 1187 1025 1 1 1207 1011 @b
-1 -1 1208 1012 1 1 1228 998 @b
-1 -1 1229 999 1 1 1249 985 @b
-1 -1 1250 986 1 1 1270 972 @b
-1 -1 1271 973 1 1 1291 959 @b
-1 -1 1292 960 1 1 1312 946 @b
-1 -1 1313 947 1 1 1333 933 @b
-1 -1 1334 934 1 1 1354 920 @b
-1 -1 1355 921 1 1 1375 907 @b
-1 -1 1376 908 1 1 1396 894 @b
-1 -1 1397 895 1 1 1417 881 @b
-1 -1 1418 882 1 1 1438 868 @b
-1 -1 1439 869 1 1 1459 855 @b
-1 -1 1460 856 1 1 1480 842 @b
-1 -1 1481 843 1 1 1501 829 @b
-1 -1 1502 830 1 1 1522 816 @b
-1 -1 1523 817 1 1 1543 803 @b
-1 -1 1544 804 1 1 1564 790 @b
-1 -1 1565 791 1 1 1584 777 @b
-1 -1 1585 778 1 1 1606 764 @b
-1 -1 1607 765 1 1 1626 751 @b
-1 -1 1627 752 1 1 1648 738 @b
-1 -1 1649 739 1 1 1668 725 @b
-1 -1 1669 726 1 1 1689 712 @b
-1 -1 1690 713 1 1 1710 699 @b
-1 -1 1711 700 1 1 1731 686 @b
-1 -1 1732 687 1 1 1752 673 @b
-1 -1 706 1323 1 1 725 1315 @b
-1 -1 726 1316 1 1 746 1308 @b
-1 -1 747 1309 1 1 767 1301 @b
-1 -1 768 1302 1 1 788 1290 @b
-1 -1 789 1291 1 1 809 1282 @b
-1 -1 810 1283 1 1 830 1272 @b
-1 -1 831 1273 1 1 851 1264 @b
-1 -1 852 1265 1 1 872 1254 @b
-1 -1 873 1255 1 1 893 1245 @b
-1 -1 894 1246 1 1 914 1235 @b
-1 -1 915 1236 1 1 935 1225 @b
-1 -1 936 1226 1 1 956 1216 @b
-1 -1 957 1217 1 1 977 1206 @b
-1 -1 978 1207 1 1 998 1197 @b
-1 -1 999 1198 1 1 1019 1187 @b
-1 -1 1020 1188 1 1 1040 1177 @b
-1 -1 1041 1178 1 1 1060 1167 @b
-1 -1 1061 1168 1 1 1082 1157 @b
-1 -1 1083 1158 1 1 1102 1147 @b
-1 -1 1103 1148 1 1 1124 1138 @b
-1 -1 1125 1139 1 1 1144 1127 @b
-1 -1 1145 1128 1 1 1165 1117 @b
-1 -1 1166 1118 1 1 1186 1107 @b
-1 -1 1187 1108 1 1 1207 1097 @b
-1 -1 1208 1098 1 1 1228 1087 @b
-1 -1 1229 1088 1 1 1249 1077 @b
-1 -1 1250 1078 1 1 1270 1067 @b
-1 -1 1271 1068 1 1 1291 1057 @b
-1 -1 1292 1058 1 1 1312 1047 @b
-1 -1 1313 1048 1 1 1333 1037 @b
-1 -1 1334 1038 1 1 1354 1026 @b
-1 -1 1355 1027 1 1 1375 1016 @b
-1 -1 1376 1017 1 1 1396 1006 @b
-1 -1 1397 1007 1 1 1417 996 @b
-1 -1 1418 997 1 1 1438 985 @b
-1 -1 1439 986 1 1 1459 976 @b
-1 -1 1460 977 1 1 1480 965 @b
-1 -1 1481 966 1 1 1501 955 @b
-1 -1 1502 956 1 1 1522 945 @b
-1 -1 1523 946 1 1 1543 935 @b
-1 -1 1544 936 1 1 1564 924 @b
-1 -1 1565 925 1 1 1584 914 @b
-1 -1 1585 915 1 1 1606 904 @b
-1 -1 1607 905 1 1 1626 893 @b
-1 -1 1627 894 1 1 1648 883 @b
-1 -1 1649 884 1 1 1668 872 @b
-1 -1 1669 873 1 1 1689 862 @b
-1 -1 1690 863 1 1 1710 851 @b
-1 -1 1711 852 1 1 1731 841 @b
-1 -1 1732 842 1 1 1752 830 @b
736 1323 -1 1 747 1322 1 736 1322 @a
-1 -1 747 1323 1 1 767 1319 @b
-1 -1 768 1320 1 1 788 1316 @b
-1 -1 789 1317 1 1 809 1312 @b
-1 -1 810 1313 1 1 830 1307 @b
-1 -1 831 1308 1 1 851 1302 @b
-1 -1 852 1303 1 1 872 1297 @b
-1 -1 873 1298 1 1 893 1292 @b
-1 -1 894 1293 1 1 914 1287 @b
-1 -1 915 1288 1 1 935 1281 @b
-1 -1 936 1282 1 1 956 1276 @b
-1 -1 957 1277 1 1 977 1269 @b
-1 -1 978 1270 1 1 998 1264 @b
-1 -1 999 1265 1 1 1019 1257 @b
-1 -1 1020 1258 1 1 1040 1251 @b
-1 -1 1041 1252 1 1 1060 1245 @b
-1 -1 1061 1246 1 1 1082 1238 @b
-1 -1 1083 1239 1 1 1102 1231 @b
-1 -1 1103 1232 1 1 1124 1224 @b
-1 -1 1125 1225 1 1 1144 1217 @b
-1 -1 1145 1218 1 1 1165 1210 @b
-1 -1 1166 1211 1 1 1186 1203 @b
-1 -1 1187 1204 1 1 1207 1196 @b
-1 -1 1208 1197 1 1 1228 1189 @b
-1 -1 1229 1190 1 1 1249 1181 @b
-1 -1 1250 1182 1 1 1270 1174 @b
-1 -1 1271 1175 1 1 1291 1166 @b
-1 -1 1292 1167 1 1 1312 1158 @b
-1 -1 1313 1159 1 1 1333 1150 @b
-1 -1 1334 1151 1 1 1354 1143 @b
-1 -1 1355 1144 1 1 1375 1135 @b
-1 -1 1376 1136 1 1 1396 1127 @b
-1 -1 1397 1128 1 1 1417 1119 @b
-1 -1 1418 1120 1 1 1438 1111 @b
-1 -1 1439 1112 1 1 1459 1102 @b
-1 -1 1460 1103 1 1 1480 1094 @b
-1 -1 1481 1095 1 1 1501 1085 @b
-1 -1 1502 1086 1 1 1522 1077 @b
-1 -1 1523 1078 1 1 1543 1069 @b
-1 -1 1544 1070 1 1 1564 1060 @b
-1 -1 1565 1061 1 1 1584 1051 @b
-1 -1 1585 1052 1 1 1606 1042 @b
-1 -1 1607 1043 1 1 1626 1033 @b
-1 -1 1627 1034 1 1 1648 1024 @b
-1 -1 1649 1025 1 1 1668 1015 @b
-1 -1 1669 1016 1 1 1689 1006 @b
-1 -1 1690 1007 1 1 1710 996 @b
-1 -1 1711 997 1 1 1731 986 @b
-1 -1 1732 987 1 1 1752 977 @b
-1 -1 1209 1323 1 1 1228 1312 @b
-1 -1 1229 1313 1 1 1249 1302 @b
-1 -1 1250 1303 1 1 1270 1295 @b
-1 -1 1271 1296 1 1 1291 1288 @b
-1 -1 1292 1289 1 1 1312 1282 @b
-1 -1 1313 1283 1 1 1333 1275 @b
-1 -1 1334 1276 1 1 1354 1269 @b
-1 -1 1355 1270 1 1 1375 1263 @b
-1 -1 1376 1264 1 1 1396 1256 @b
-1 -1 1397 1257 1 1 1417 1251 @b
-1 -1 1418 1252 1 1 1438 1244 @b
-1 -1 1439 1245 1 1 1459 1238 @b
-1 -1 1460 1239 1 1 1480 1230 @b
-1 -1 1481 1231 1 1 1501 1224 @b
-1 -1 1502 1225 1 1 1522 1217 @b
-1 -1 1523 1218 1 1 1543 1210 @b
-1 -1 1544 1211 1 1 1564 1203 @b
-1 -1 1565 1204 1 1 1584 1196 @b
-1 -1 1585 1197 1 1 1606 1189 @b
-1 -1 1607 1190 1 1 1626 1182 @b
-1 -1 1627 1183 1 1 1648 1174 @b
-1 -1 1649 1175 1 1 1668 1167 @b
-1 -1 1669 1168 1 1 1689 1159 @b
-1 -1 1690 1160 1 1 1710 1151 @b
-1 -1 1711 1152 1 1 1731 1143 @b
-1 -1 1732 1144 1 1 1752 1135 @b
gR
gS 390 309 1572 1170 rC
704.5 1277.5 14.5 @i
1 lw
704.5 1277.5 14 @e
704.5 1186.5 14.5 @i
704.5 1186.5 14 @e
704.5 1121.5 14.5 @i
704.5 1121.5 14 @e
704.5 1076.5 14.5 @i
704.5 1076.5 14 @e
704.5 1011.5 14.5 @i
704.5 1011.5 14 @e
704.5 952.5 14.5 @i
704.5 952.5 14 @e
903.5 1232.5 14.5 @i
903.5 1232.5 14 @e
903.5 1108.5 14.5 @i
903.5 1108.5 14 @e
903.5 1076.5 14.5 @i
903.5 1076.5 14 @e
903.5 1011.5 14.5 @i
903.5 1011.5 14 @e
1448.5 764.5 14.5 @i
1448.5 764.5 14 @e
30 29 1606 887.5 @j
29 28 1606 887.5 @f
1626.5 796.5 14.5 @i
1626.5 796.5 14 @e
1668.5 861.5 14.5 @i
1668.5 861.5 14 @e
1721.5 796.5 14.5 @i
1721.5 796.5 14 @e
1721.5 673.5 14.5 @i
1721.5 673.5 14 @e
1752.5 738.5 14.5 @i
1752.5 738.5 14 @e
1752.5 640.5 14.5 @i
1752.5 640.5 14 @e
0 0 1 :F
679 1323 -1 1 705 1322 1 679 1322 @a
679 1194 -1 1 705 1193 1 679 1193 @a
679 1064 -1 1 705 1063 1 679 1063 @a
679 934 -1 1 705 933 1 679 933 @a
679 804 -1 1 705 803 1 679 803 @a
679 674 -1 1 705 673 1 679 673 @a
679 544 -1 1 705 543 1 679 543 @a
1752 1323 -1 1 1778 1322 1 1752 1322 @a
1752 1194 -1 1 1778 1193 1 1752 1193 @a
1752 1064 -1 1 1778 1063 1 1752 1063 @a
1752 934 -1 1 1778 933 1 1752 933 @a
1752 804 -1 1 1778 803 1 1752 803 @a
1752 674 -1 1 1778 673 1 1752 673 @a
1752 544 -1 1 1778 543 1 1752 543 @a
-1 -1 705 1348 1 1 704 1322 @b
-1 -1 915 1348 1 1 914 1322 @b
-1 -1 1125 1348 1 1 1124 1322 @b
-1 -1 1334 1348 1 1 1333 1322 @b
-1 -1 1544 1348 1 1 1543 1322 @b
-1 -1 1753 1348 1 1 1752 1322 @b
-1 -1 705 544 1 1 704 518 @b
-1 -1 915 544 1 1 914 518 @b
-1 -1 1125 544 1 1 1124 518 @b
-1 -1 1334 544 1 1 1333 518 @b
-1 -1 1544 544 1 1 1543 518 @b
-1 -1 1753 544 1 1 1752 518 @b
-1 -1 705 1323 1 1 704 543 @b
704 1323 -1 1 1753 1322 1 704 1322 @a
-1 -1 1753 1323 1 1 1752 543 @b
704 544 -1 1 1753 543 1 704 543 @a
1 G
-8 -8 402 321 8 8 394 313 @b
gS
4.17 4.163 scale
151.815 322.575 :T
-151.815 -322.575 :T
151.815 322.575 :M
0 G
f1_12 sf
(0)S
141.022 291.59 :M
.829(0.2)A
141.022 259.645 :M
.829(0.4)A
141.022 228.661 :M
.829(0.6)A
141.022 197.676 :M
.829(0.8)A
151.815 166.692 :M
(1)S
141.022 135.707 :M
.829(1.2)A
165.006 338.667 :M
(0)S
gR
gR
1 G
gS 0 0 2300 3042 rC
gS
4.17 4.163 scale
118.958 235.626 :T
270 rotate
-118.958 -235.626 :T
118.958 235.626 :M
0 G
f1_12 sf
-.791(Gap)A
gR
390 309 1572 1170 rC
gS
4.17 4.163 scale
210.814 338.667 :T
-210.814 -338.667 :T
210.814 338.667 :M
0 G
f1_12 sf
.829(0.2)A
260.939 338.667 :M
.829(0.4)A
310.824 338.667 :M
.829(0.6)A
360.95 338.667 :M
.829(0.8)A
416.831 338.667 :M
(1)S
427.863 167.652 :M
.829(0.0)A
427.863 205.602 :M
.829(0.1)A
427.863 241.631 :M
.829(0.3)A
272.931 352.598 :M
.873(cos\(2)A
306.987 352.598 :M
f5_12 sf
.358(f)A
f1_12 sf
(\))S
265.016 121.536 :M
f4_14 sf
2.131(d-wave)A
427.863 278.62 :M
f1_12 sf
.829(0.6)A
endp
end		